\documentclass[epj]{svjour}
\usepackage{graphics}

\usepackage{marginnote}

\usepackage{epsfig}
\usepackage{graphicx}
\usepackage{mathptmx}      
\usepackage{latexsym}
\usepackage{bm}
\usepackage{color}
\usepackage{subfigure}
\usepackage{amsmath}

\usepackage{stackrel}

\def\slashchar#1{\setbox0=\hbox{$#1$}
   \dimen0=\wd0 \setbox1=\hbox{/} \dimen1=\wd1
   \ifdim\dimen0>\dimen1 \rlap{\hbox to \dimen0{\hfil/\hfil}} #1
   \else  \rlap{\hbox to \dimen1{\hfil$#1$\hfil}} / \fi}

\newcommand{\ignore}[1]{}

\newcommand{\KG}{{\textrm{\scriptsize KG}}}
\newcommand{\NR}{{\textrm{\scriptsize NR}}}

\begin{document}
\title{Nonlinear Klein-Gordon equation and the Bose-Einstein condensation}
\author{E.~Meg\'{\i}as\inst{1} \and M.J.~Teixeira\inst{2} \and V.S.~Timoteo\inst{3} \and A.~Deppman\inst{2}
}                     
%
%
\institute{Departamento de F{\'\i}sica At\'omica, Molecular y Nuclear and
  Instituto Carlos I de F{\'\i}sica Te\'orica y Computacional, Universidad
  de Granada, E-18071 Granada, Spain
  \and
  Instituto de F\'isica, Universidade de S\~ao Paulo, S\~ao Paulo, SP 05508-090, Brasil
 \and  
  Grupo de \'Optica e Modelagem Num\'erica, Faculdade de Tecnologia GOMNI/FT - Universidade Estadual de Campinas - UNICAMP, Limeira, SP 13484-332, Brasil
}%
\date{Received: date / Revised version: date}
%
\abstract{
The interest in the Klein-Gordon equation with different potentials has increased in recent years due to its possible applications in Cosmology, Hadron Physics and High-Energy Physics. In this work we investigate the solutions of the Klein-Gordon equation for bosons under the influence of an external potential by using the Feshbach-Villars method. We present detailed results for two cases: the Coulombic potential and the harmonic potential. For the latter case, we studied the effects of self-interacting particles by adopting a mean-field approach. We show that our results converge smoothly to the solution of the Schr\"odinger equation for the same systems as the relativistic effects diminish.
\PACS{
      {PACS-key}{discribing text of that key}   \and
      {PACS-key}{discribing text of that key}
     } 
} 
\maketitle

\noindent{\it Introduction:} The Klein-Gordon (KG) equation has been investigated by various methods due to its application in different fields. The case of a scalar field has an increasing interest because of its connections with the axion, a hypothesized particle that may have a relevant role in Astrophysics and in studies of Dark-Matter~\cite{Hogan,Frieberg,EbyLeembruggen,EbyMadelyn}.

In Astrophysics, the investigations of a possible bosonic phase in massive stars has triggered the study of the solutions of the relativistic equation with an external potential. The possibility of a condensate regime inside the stars has been considered. Several methods have been used to solve the equation for different potentials~\cite{GrundlandInfeld,EbyStreet}. In the case of self-interacting particles, the problem becomes particularly difficult, and in Ref.~\cite{EbySuranyi} the authors claim that it is not possible to find a solution for this case. Nevertheless, they investigated the axion star and the possibility of the formation of a condensate.

Systems formed by interacting bosons are also of interest in High-Energy Physics and in Hadron Physics, where the possibility of a gluonic field in the condensate regime has been considered~\cite{Rezaeian,Jankowski,Fujii}. With the advent of colliders such as the EIC, FAIR and NICA, with the capability to investigate the hadron structure in details, we can expect important advances in our understanding of the complex hadronic structure. The effects of the interaction of the components of the system with an external potential has been extensively studied in connection with the Bose-Einstein condensate~\cite{Ketterle-vanDruten-PRL-1996,Bagnato-Pritchard-Kleppner}. The interaction can favour or disfavour the formation of the condensate. The correlations and interactions among the particles do not prevent the formation of the condensate~\cite{RepulsiveInteractionBEC}, and can lead to non-extensive forms of the condensate described in terms of the Tsallis statistics~\cite{Plastino-Tsallis-2016,Guerrero-Gonzalez}. The characteristics of the non extensive condensate has been described in Ref.~\cite{Megias-Varese-Gammal-Deppman-2021,Rajagopal-Lenzi}.

In this work, we investigate the solutions of the KG equation for a system of bosons under the influence of an external potential, in particular for the Coulombic potential and for the harmonic potential, and then we include the self-interaction through a mean-field approach. We investigate the ground-state solution and obtain the chemical potential for the case of the external harmonic potential, and with the self-interaction regulated by the scattering length. We use the Feshbach-Villars formalism, which allows us to evaluate the transition from the relativistic to the non-relativistic regimes. This is particularly useful in the case of self-interacting particles in a harmonic potential, for which the Schr\"odinger equation is known~\cite{Gammal:1999,Adhikari:2000}. We show that our results with the KG equation smoothly approach those for the Schorödinger equation as the systems become non-relativistic.

~

\noindent {\it The Feshbach-Villars formalism: }
The KG equation for free particles is
\begin{equation}
  \left[ \frac{1}{c^2} \frac{\partial}{\partial t^2} - \vec{\nabla}^2 + \frac{m^2 c^2}{\hbar^2} \right] \psi(t,\vec{r})  = 0\,. \label{eq:KG_massless}
\end{equation}
This equation can be written in a convenient form by using the Feshbach-Villars formalism~\cite{Feshbach:1958wv} (see also Ref.~\cite{Alberto:2017pkj}). In this case, we define the spinor
\begin{equation}
\Psi(t,\vec{r}) = \left(
\begin{array}{c}
\varphi(t,\vec{r}) \\
\chi(t,\vec{r})    
\end{array}  \right)\,,  
\end{equation}
where $\varphi(t,\vec{r})$ and $\chi(t,\vec{r})$ are wave-functions related to $\psi(t,\vec{r})$ by
\begin{equation}
 \begin{cases}
   \psi(t,\vec{r}) &= \quad \varphi(t,\vec{r}) + \chi(t,\vec{r}) \\
   i \hbar \frac{\partial \psi(t,\vec{r})}{dt} &= \quad mc^2 \left( \varphi(t,\vec{r}) - \chi(t,\vec{r})  \right)
 \end{cases} \,.
\end{equation}
In the Feshbach-Villars formalism, the KG equation is written for the spinor as a Schr\"odinger-like equation, i.e.
\begin{equation}
i\hbar \frac{\partial}{\partial t} \Psi(t,\vec{r}) = H \Psi(t,\vec{r}) \,,  \label{eq:dt_H}
\end{equation}
where
\begin{equation}
 H = (\sigma_3 + i \sigma_2) \frac{\hat{\vec{p}}^2}{2m} + m c^2 \sigma_3 \,.
\end{equation}
In the equation above $\sigma_i \; (i=1,2,3)$ are the Pauli matrices, and $\hat{\vec{p}} = -i \hbar \vec{\nabla}$. The conserved charge and current are given by
\begin{equation}
 \begin{cases}
\rho = \Psi^\dagger \sigma_3 \Psi  \\
\vec{J} = \frac{\hbar}{i2m} \left[ \Psi^\dagger \sigma_3 (\sigma_3 + i \sigma_2) \vec{\nabla} \Psi - (\vec{\nabla} \Psi^\dagger) \sigma_3 (\sigma_3 + i \sigma_2) \Psi \right] 
 \end{cases} \,,
\end{equation}
respectively. The spinor representation is equivalent to the coupled differential equations
\begin{equation}
 \begin{cases}
i\hbar \frac{\partial \varphi}{\partial t} = \frac{\hat{\vec{p}}^2}{2m}(\varphi + \chi) + m c^2 \varphi 
\\
i\hbar \frac{\partial \chi}{\partial t} = -\frac{\hat{\vec{p}}^2}{2m}(\varphi + \chi) - m c^2 \chi \label{eq:varphi-chi}
 \end{cases} \,.
\end{equation}
The Feshbach-Villars formalism becomes especially convenient to study the non-relativistic regime, in which the relativistic wave equation reduces to the solution of the Schr\"odinger equation. We will study below some solutions of the KG equation for different potentials. 

The time-independent form of the KG equation is
\begin{equation}
H \Psi(\vec{r}) = E \Psi(\vec{r}) \,,
\end{equation}
and the corresponding solution of the time-dependent equation, Eq.~(\ref{eq:dt_H}), is
\begin{equation}
\Psi(t,\vec{r}) = e^{-iEt/\hbar} \Psi(\vec{r}) \,.
\end{equation}

~

\noindent{\it Introduction of the chemical potential:} In the present work we are interested in the investigation of the Bose-Einstein condensate formed in a relativistic system. We include the chemical potential, by assuming that the ground-state energy coincides with the chemical potential. The stationary-state solution is
\begin{equation}
    \Psi^{\pm}(t,\vec{r}) = e^{\mp i\mu t/\hbar} \Psi(\vec{r}) \,, \label{eq:psi_mu}
\end{equation}
where $\pm$ corresponds to the positive(negative) chemical potential solution. Then, the original KG equation, Eq.~(\ref{eq:KG_massless}), leads to a time-independent KG equation which writes
\begin{equation}
\left[ - \hbar^2 c^2 \vec{\nabla}^2   +  m^2 c^4 - \mu^2 \right] \psi({\vec{r}}) = 0 \,, \label{eq:KG_time_indep}
\end{equation}
whose solution in the spinor representation is
\begin{equation}
\Psi_{\vec{p}}^\pm(t,\vec{r}) =  e^{ \mp i \mu t/\hbar} \Psi_0^\pm(\vec{p}) \left(  A_\pm e^{i\vec{p} \cdot \vec{r} /\hbar} + B_\pm  e^{-i\vec{p} \cdot \vec{r} /\hbar} \right)\,,
\end{equation}
where the spinor
\begin{equation}
\Psi_0^\pm(\vec{p}) = \left(
\begin{array}{c}
\varphi_0^\pm(\vec{p}) \\
\chi_0^\pm(\vec{p})    
\end{array}  \right)\,, 
\end{equation}
has the amplitudes
\begin{equation}
 \begin{cases}
 \varphi_0^\pm(\vec{p}) = \frac{\pm \mu + mc^2}{2\sqrt{mc^2 \mu}}  \\   \chi_0^\pm(\vec{p}) = \frac{\mp \mu + mc^2}{2\sqrt{mc^2 \mu}}  
 \end{cases} \,.
\end{equation}
Notice that $[\varphi_0^\pm(\vec{p})]^2 - [\chi_0^\pm(\vec{p})]^2 = \pm 1$, and  $\mu \equiv E_p = \sqrt{p^2 c^2 + m^2 c^4}$.

~

\begin{figure}[t]
 \begin{subfigure}{}
  \includegraphics[width=0.27\textwidth]{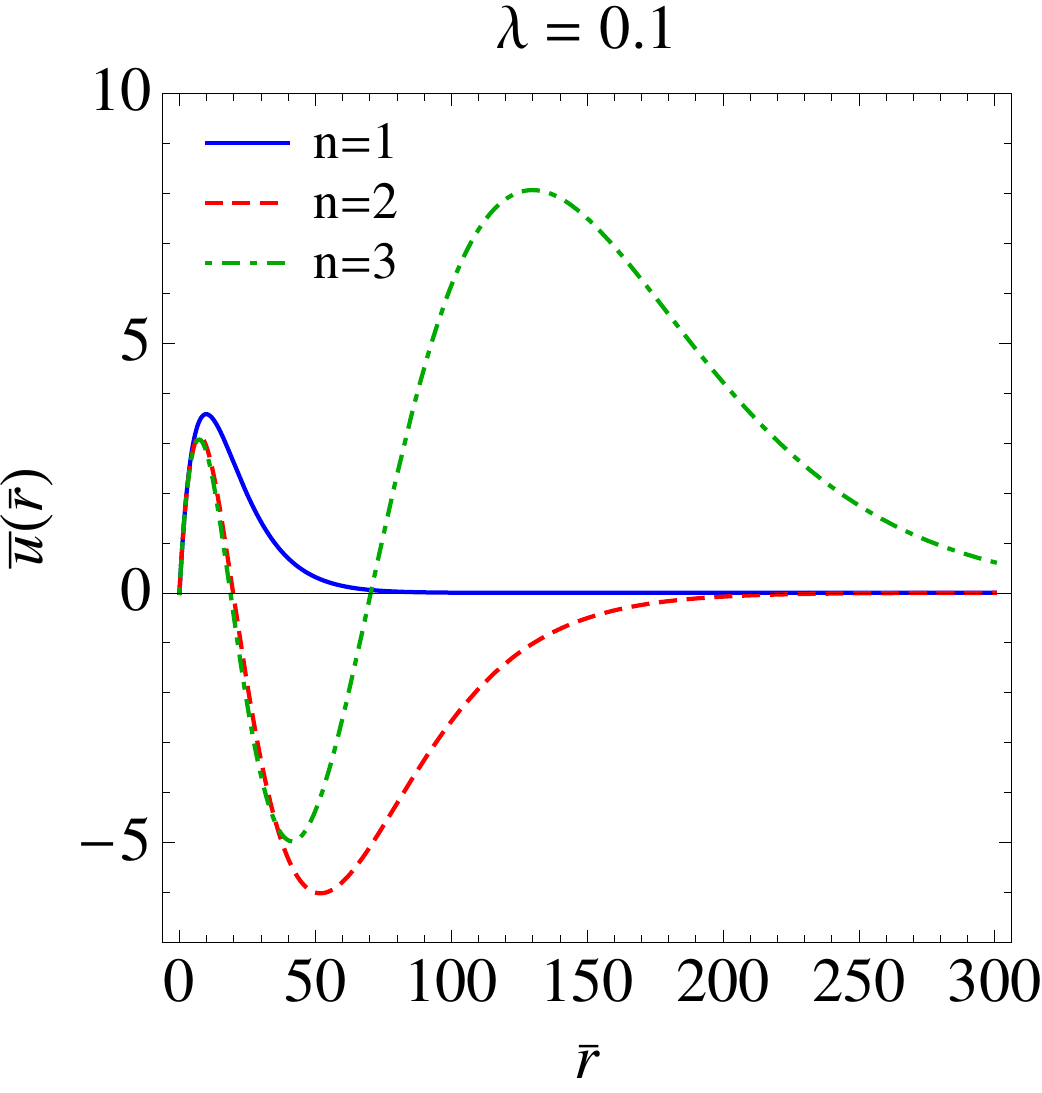} 
 \end{subfigure} \hspace{0.3cm}
  \begin{subfigure}{}
  \includegraphics[width=0.27\textwidth]{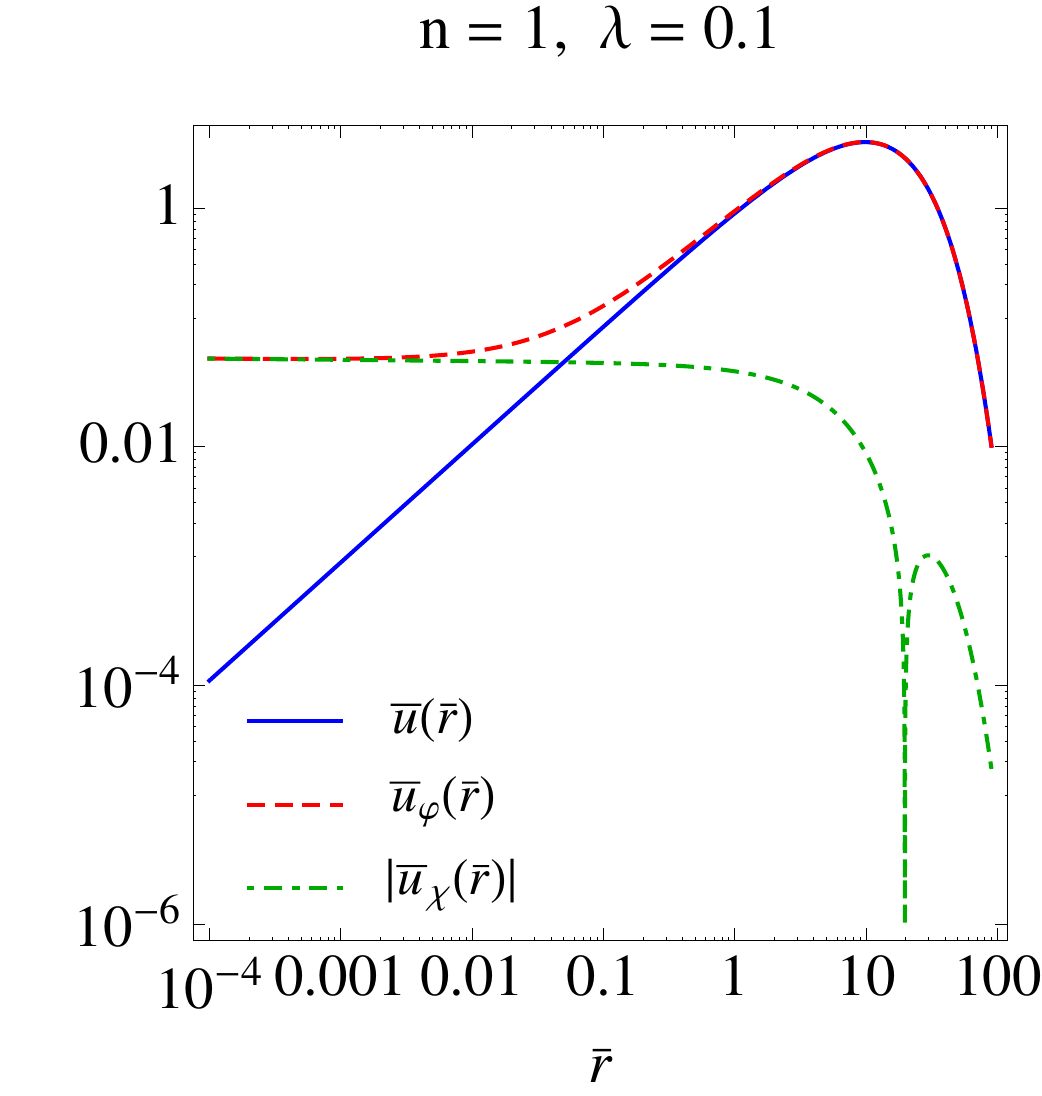} 
 \end{subfigure} \hspace{0.3cm}
 \begin{subfigure}{}
  \includegraphics[width=0.29\textwidth]{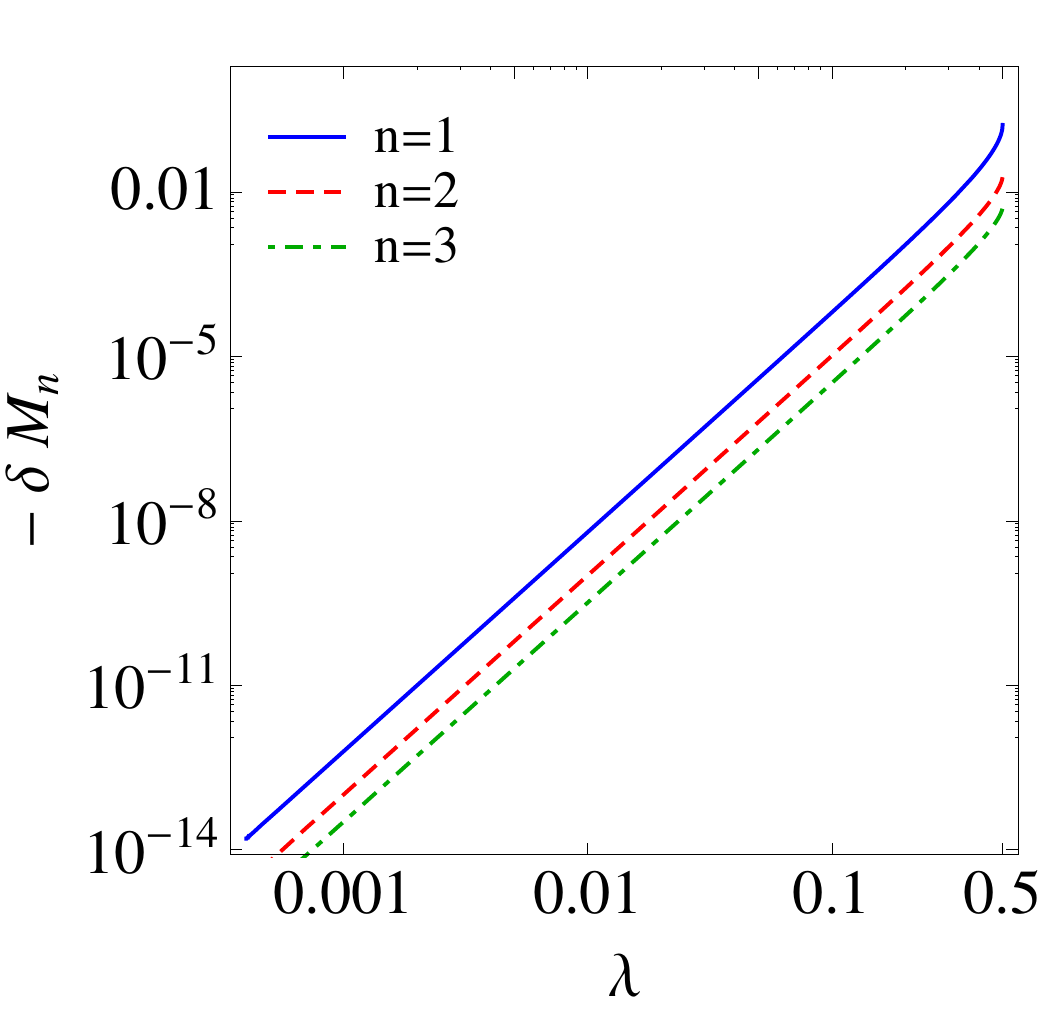} 
 \end{subfigure} 
 \caption{Left panel: Dimensionless reduced wave functions $\bar u(\bar r)$ obtained by solving the KG equation with a Coulombic potential (cf. Eqs.~(\ref{eq:ur_Coulomb}) and (\ref{eq:Coulomb_dimensionless})) We display the wave functions for the first three levels: $n = 1, 2, 3$. Middle panel: Reduced wave function $\bar u(\bar r)$ and the spinor components ($\bar u_\varphi(\bar r)$, $\bar u_\chi(\bar r)$) for $n = 1$. In both panels we have considered $\lambda = 0.1$ and we have normalized the wave functions such that $\bar u(\bar r) \stackrel[r \to 0]{\simeq}{} \bar r^{\beta_+}$ (cf. Eq.~(\ref{eq:u_r0})). Right panel: $-\delta M_n$ as a function of $\lambda$ for the first three levels (cf. Eq.~(\ref{eq:MKG_c_large})).}
 \label{fig:VCoulomb}
\end{figure}

\noindent {\it Introduction of an external potential:} The Bose-Einstein condensation can happen even in the presence of an external potential. To study the effects of an external potential, let us solve the time-independent KG equation 
\begin{equation}
\vec{\nabla}^2 \psi(\vec{r}) + \frac{1}{\hbar^2 c^2} \left[ (\mu-V(\vec{r}))^2 - m^2 c^4 \right] \psi(\vec{r}) = 0  \,, \label{eq:Psi_V}
\end{equation}
in different cases. This equation has been obtained and studied in the literature in several contexts, see e.g. Refs.~\cite{Dominguez-Adame:1989gml,Chen:2005,Zhang:2005,Chen:2005plA,ChenLu:2006,Zhang:2006,Ducharme:2010,Sharma:2011,Bi:2012rx}. It can be derived by applying the correspondence principle in quantum mechanics: $E \to i\hbar \frac{\partial}{\partial t}$, $\vec{p} \to -i\hbar \vec{\nabla}$; to the relativistic expression for the energy $E = \sqrt{ \vec{p}^2 c^2 + m^2 c^4} + V(\vec{r})$. Using the Feshbach-Villars formalism we obtain the relation
\begin{equation}
     \mu \psi = mc^2 (\varphi - \chi) + V(\vec{r}) (\varphi + \chi) \,, \label{eq:psi_phichimu}
\end{equation}
with $\psi = \varphi + \chi $. Then, we find from Eq.~(\ref{eq:psi_phichimu}) that the spinor components $\varphi$ and $\chi$ can be computed from the wave function $\psi$ as follows
\begin{equation}
 \begin{cases}
    \varphi(\vec{r}) = \frac{1}{2}\left( 1 + \frac{\mu - V(\vec{r})}{mc^2} \right) \psi(\vec{r}) \\
    \chi(\vec{r}) = \frac{1}{2}\left( 1 - \frac{\mu - V(\vec{r})}{mc^2} \right) \psi(\vec{r})  \label{eq:phichi}
 \end{cases} \,.
\end{equation}
In the massless case, one has to consider $m=0$ in Eq.~(\ref{eq:Psi_V}), and replace $m \to \mathcal N$ in Eqs.~(\ref{eq:psi_phichimu}) and (\ref{eq:phichi}), where $\mathcal N$ is an arbitrary real nonzero parameter~\cite{Silenko:2019brp}.

The Schr\"odinger equation can be obtained from the KG equation~(\ref{eq:Psi_V}) by considering the non-relativistic limit, i.e. $c \to \infty$. To see that, let us define the energy with the rest-mass energy subtracted, $\bar\mu$, as
\begin{equation}
\mu \equiv mc^2 + \bar\mu \,. \label{eq:mu_mub}
\end{equation}
Then, we can write the KG equation~(\ref{eq:Psi_V}) in a form similar to the Schr\"odinger equation, but with an explicit relativistic correction. This is
\begin{equation}
-\frac{\hbar^2}{2m}\vec{\nabla}^2 \psi(\vec{r}) +  \left( V(\vec{r}) - \bar\mu   \right) \left[ 1 - \frac{(V(\vec{r}) - \bar\mu)}{2mc^2} \right] \psi(\vec{r}) = 0 \,, \label{eq:Psi_V3}
\end{equation}
which is completely equivalent to Eq.~(\ref{eq:Psi_V}). Then the Schr\"odinger equation is obtained from the KG equation in the limit $|V(\vec{r}) - \bar\mu|  \ll  m c^2$. Notice that in this limit one has from Eq.~(\ref{eq:phichi}) that $\varphi(\vec{r}) \stackrel[c\to \infty]{\simeq}{} \psi(\vec{r})$ and $\chi(\vec{r}) \stackrel[c\to \infty]{\simeq}{} 0$. 

~

\noindent{\it Coulombic potential:} Let us study the case of a Coulombic potential,
\begin{equation}
V(r) = -\frac{g}{r} \,, \qquad g > 0 \,. \label{eq:V_Coulombic}
\end{equation}
We consider Eq.~(\ref{eq:Psi_V}) in spherical coordinates, and assume $\ell = 0$ (zero angular momentum), obtaining the equation
\begin{equation}
\psi^{\prime\prime}(r) + \frac{2}{r} \psi^\prime(r) + \frac{1}{\hbar^2 c^2}\left[ (\mu - V(r))^2 - m^2 c^4 \right] \psi(r) = 0 \,.  \label{eq:KG_spherical}
\end{equation}
This equation can be written in a simpler form by introducing the reduced radial wave function
\begin{equation}
    u(r) \equiv r \psi(r) \,.
\end{equation}
Then one gets
\begin{equation}
u^{\prime\prime}(r)    + \frac{1}{\hbar^2 c^2} \left[ (\mu - V(r))^2 - m^2 c^4 \right] u(r) = 0 \,.  \label{eq:u}
\end{equation}
The solution of Eq.~(\ref{eq:u}) with the potential of Eq.~(\ref{eq:V_Coulombic}) can be obtained analytically as a linear combination of two independent solutions, i.e.
\begin{equation}
u(r) =     r^{\beta_+} e^{-\frac{\Delta}{\hbar c} r } \left[ C_1 \cdot L_{-\beta_+ + \frac{g \mu}{\hbar c \Delta}}^{\beta_+ - \beta_-}\left( \frac{2\Delta}{\hbar c} r \right)  + C_2 \cdot U\left(\beta_+ - \frac{g\mu}{\hbar c \Delta}, 2 \beta_p, \frac{2\Delta }{\hbar c} r \right) \right] \,, \label{eq:ur_Coulomb}
\end{equation}
where $L_{n}^a(z)$ and $U(a,b,z)$ are the generalized Laguerre polynomial and the confluent hypergeometric function, respectively, while we have defined $\Delta \equiv \sqrt{m^2 c^4 - \mu^2}$ and
\begin{equation}
 \beta_\pm \equiv \frac{1}{2}\left( 1 \pm \sqrt{1-\frac{4 g^2}{\hbar^2 c^2 }} \right) \,.
\end{equation}
The solution at short distances behaves as
\begin{equation}
u(r) \stackrel[r \to 0]{\simeq}{} C_- r^{\beta_-} + C_+ r^{\beta_+} \,,  \label{eq:u_r0}
\end{equation}
with $+(-)$ standing for the dominant behavior of the $L_n^a(z)(U(a,b,z))$ solution. The physical (regular) solution corresponds to $r^{\beta_+}$, so that the coefficient $C_-$ should vanish and then we should consider $C_2=0$ in Eq.~(\ref{eq:ur_Coulomb}). From the behavior of the $L$-solution at large distances, one finds that the square-integrable condition for the wave function leads to the exact result for the eigenvalue problem of Eq.~(\ref{eq:u}) which we will denote by $\mu_n^{\KG}$, corresponding to the solutions of the equation
\begin{equation}
\frac{g\mu_n^{\KG}}{\hbar c \sqrt{m^2 c^4 - \mu_n^{\KG \, 2} } } - \beta_{+}   = n-1  \,, \qquad n = 1,2,3, \cdots \,,
\end{equation}
leading to the chemical potential
\begin{equation}
\mu_n^{\KG} = m c^2 \frac{n - 1 +\beta_{+}}{\sqrt{\frac{g^2}{(\hbar c)^2} + (n - 1 +\beta_{+})^2}} \,, \qquad n = 1,2,3,\cdots \,. \label{eq:munKG}
\end{equation}
Notice that the form of Eq.~(\ref{eq:munKG}) implies that $|\mu_n^{\KG}| < mc^2$, so that the eigenfunction associated to the eigenvalue $\mu_n^{\KG}$ behaves as $u(r) \stackrel[r \to \infty]{\sim}{} e^{-\kappa r}$ with $\kappa \equiv \Delta_n/(\hbar c) > 0$, thus corresponding to a bound state.  To consider the non-relativistic limit, we make $c$ to be large enough. Then we have
\begin{equation}
\beta_\pm =   \left\{ 
\begin{array}{cc}
1 - \frac{1}{(\hbar c)^2} g^2 + \cdots    & \hspace{0.8cm}   (+)  \\
\frac{1}{(\hbar c)^2} g^2 + \cdots &  \qquad\hspace{0.1cm}   (-)  
\end{array} \,. \right. \label{eq:betapm}
\end{equation}
We can compare our results in the non-relativistic limit with the results obtained with the Schr\"odinger equation. In this latter case, the eigenvalues are
\begin{equation}
\bar \mu_{n}^{\NR} = - \frac{m g^2}{2\hbar^2 n^2} \,, \qquad n = 1,2,3, \cdots \,.     \label{eq:mun}
\end{equation}
Let us define $\delta \mu_n$ as the difference between the result for the eigenvalue obtained with the KG equation (after subtracting the rest-mass energy), and the non-relativistic  result of Eq.~(\ref{eq:mun}), i.e.
\begin{equation}
   \mu_n^{\KG} = m c^2 + \bar \mu_n^{\NR} + \delta \mu_n \,. \label{eq:mnKG}
\end{equation}
In the limit of $c \gg 1$, the expression of Eq.~(\ref{eq:munKG}) has the following expansion
\begin{equation}
\mu_n^{\KG} \stackrel[c \gg 1]{=}{}     m c^2 \left[ 1 -\frac{g^2}{2(\hbar c)^2 n^2 } - \frac{g^4}{(\hbar c)^4} \left( \frac{1}{n^3} - \frac{3}{8} \frac{1}{n^4} \right) + {\mathcal O}\left( c^{-6} \right) \right] \,, \label{eq:muKG_c_large}
\end{equation}
so that one can easily identify the correction $\delta\mu_n$ at order ${\mathcal O}(c^{-2})$ from a comparison with Eq.~(\ref{eq:mnKG}). Before displaying the results, it is convenient to define the dimensionless variables
\begin{equation}
\bar r \equiv \frac{mc^2}{\hbar c} r \,, \qquad \bar u(\bar r) \equiv \sqrt{\frac{\hbar c}{mc^2}} \cdot u(r) \,, \qquad M_n^{\KG} \equiv \frac{\mu_n^{\KG}}{m c^2} \,, \qquad \lambda \equiv \frac{g}{\hbar c} \,.     \label{eq:Coulomb_dimensionless}
\end{equation}
Then, the dimensionless eigenvalues 
\begin{equation}
M_n^{\KG} = \frac{n-1+\beta_+}{\sqrt{\lambda^2 + (n-1+\beta_+)^2}} \qquad \textrm{with} \qquad \beta_{\pm} = \frac{1}{2}\left( 1 \pm \sqrt{1-4\lambda^2} \right)
\end{equation}
are functions only of the parameter $\lambda$. In particular, the decomposition of Eq.~(\ref{eq:mnKG}) writes in dimensionless units as
\begin{equation}
M_n^{\KG} = 1 - \frac{\lambda^2}{2n^2} + \delta M_n  \quad \textrm{with} \quad \delta M_n \equiv \frac{\delta \mu_n}{mc^2} = -\lambda^4 \left( \frac{1}{n^3} - \frac{3}{8} \frac{1}{n^4}\right) + \mathcal O\left( \lambda^6 \right)   \,.    \label{eq:MKG_c_large}
\end{equation}
The non-relativistic limit corresponds to $\lambda \to 0$, or equivalently $c \to \infty$. We display in Fig.~\ref{fig:VCoulomb} (left and middle panels) the dimensionless reduced wave function $\bar u(\bar r)$, as well as the spinor components $\bar u_\varphi(\bar r) \equiv \sqrt{\frac{\hbar c}{mc^2} } \cdot r \varphi(r)$ and  $\bar u_\chi(\bar r) \equiv \sqrt{\frac{\hbar c}{mc^2} } \cdot r \chi(r)$. We display also in Fig.~\ref{fig:VCoulomb} (right) the behavior of $\delta M_n$ as a function of~$\lambda$. Notice that $\delta M_n \stackrel[\lambda \to 0]{\longrightarrow}{} 0$, as it is already clear from Eq.~(\ref{eq:MKG_c_large}), showing that the result of the Schr\"odinger equation is recovered in the non-relativistic limit. 
\begin{figure}[t]
 \begin{subfigure}{}
  \includegraphics[width=0.28\textwidth]{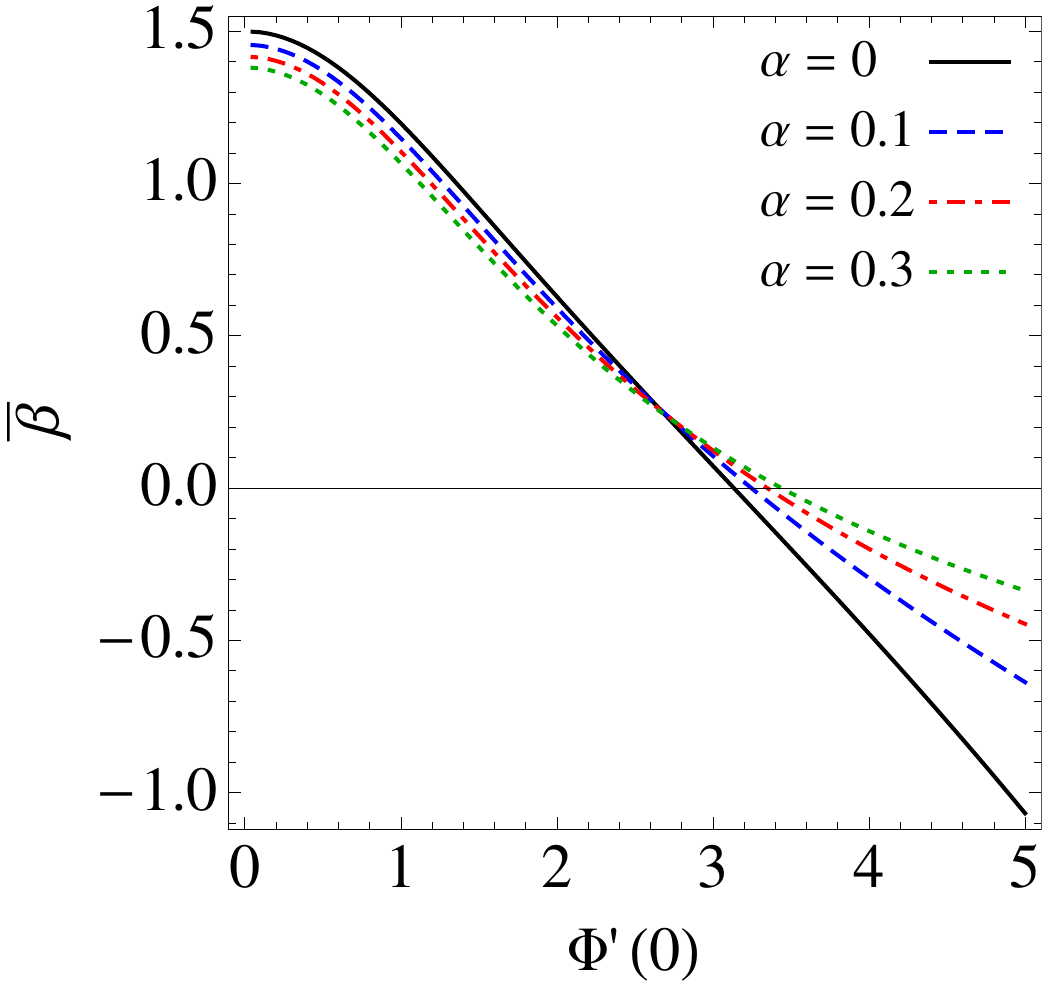} 
 \end{subfigure}  \hspace{0.3cm}
 \begin{subfigure}{}
  \includegraphics[width=0.29\textwidth]{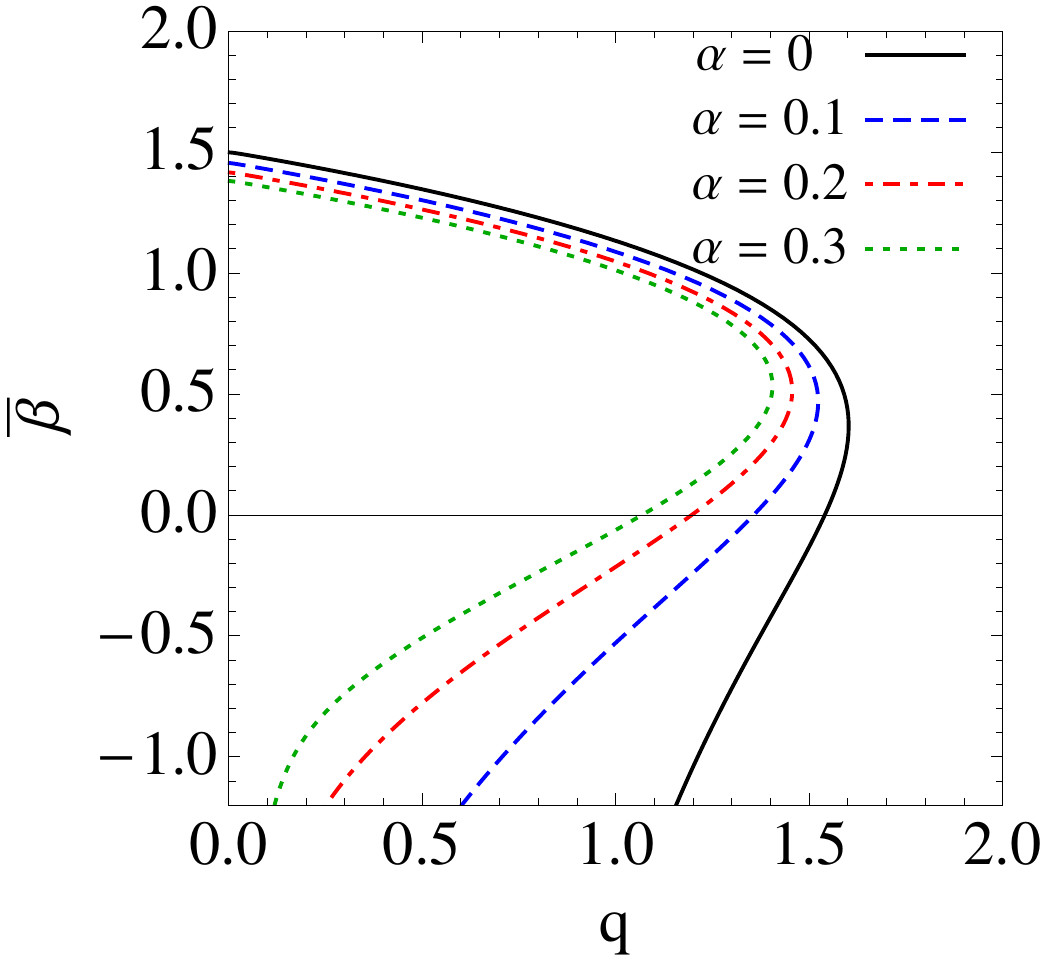} 
 \end{subfigure} \hspace{0.3cm}
  \begin{subfigure}{}
  \includegraphics[width=0.26\textwidth]{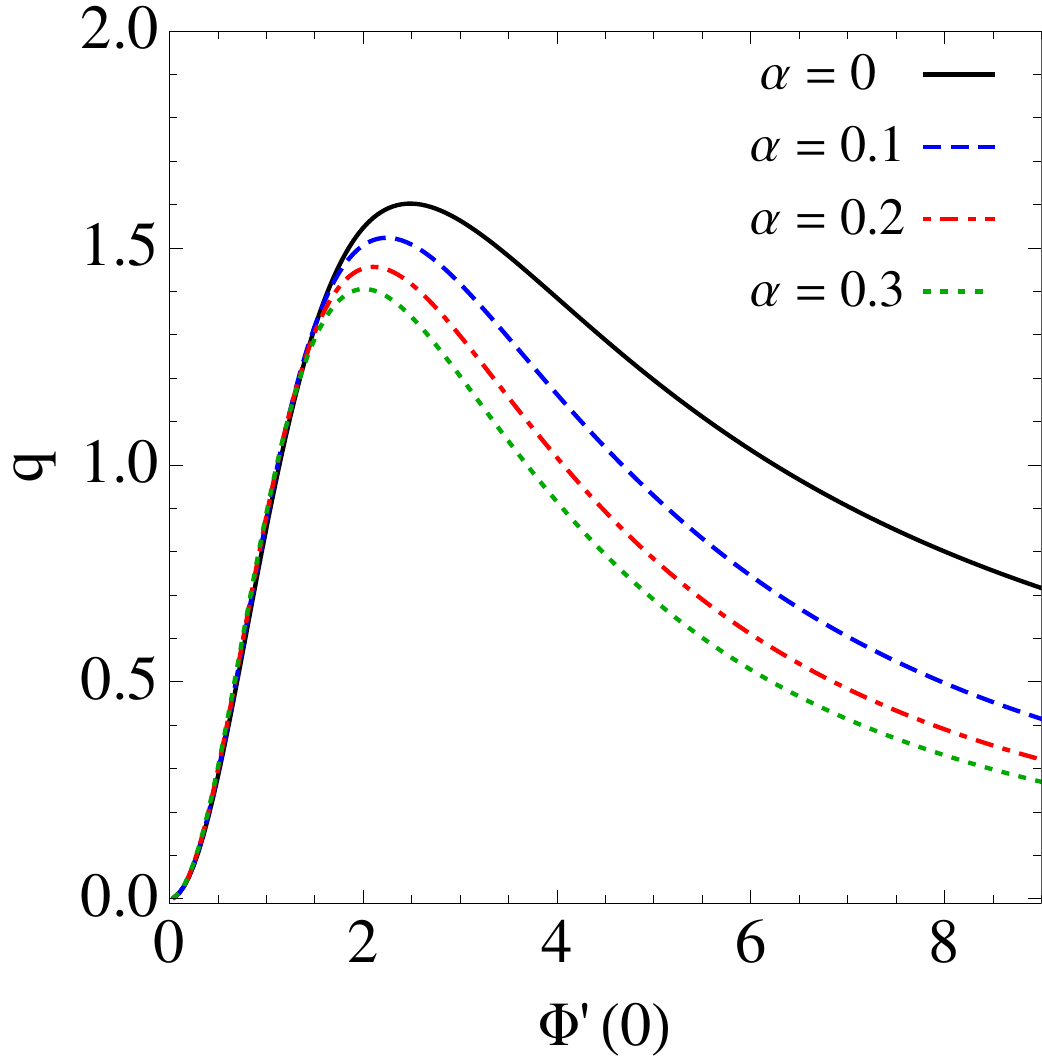} 
 \end{subfigure}
 \caption{Left and middle panels: Results for the dimensionless chemical potential $\bar\beta$ after solving the Gross-Pitaevskii nonlinear equation in the relativistic case, Eq.~(\ref{eq:KG_Pitaevskii2}). We display the relativistic result of $\bar \beta$ as a function of $\Phi^\prime(0)$ (left panel) and as a function of $q$ (middle panel) for $\alpha = 0.1, 0.2$ and $0.3$, and compared to the non-relativistic result corresponding to $\alpha = 0$ and displayed as solid (black) line. Right panel: $q$ as a function of $\Phi^\prime(0)$ for $\alpha = 0, 0.1, 0.2$ and $0.3$. Notice that $q(\alpha = 0) = n$. 
 }
 \label{fig:Pitaevskii_rel}
\end{figure}

\begin{figure}[t]
 \begin{subfigure}{}
  \includegraphics[width=0.28\textwidth]{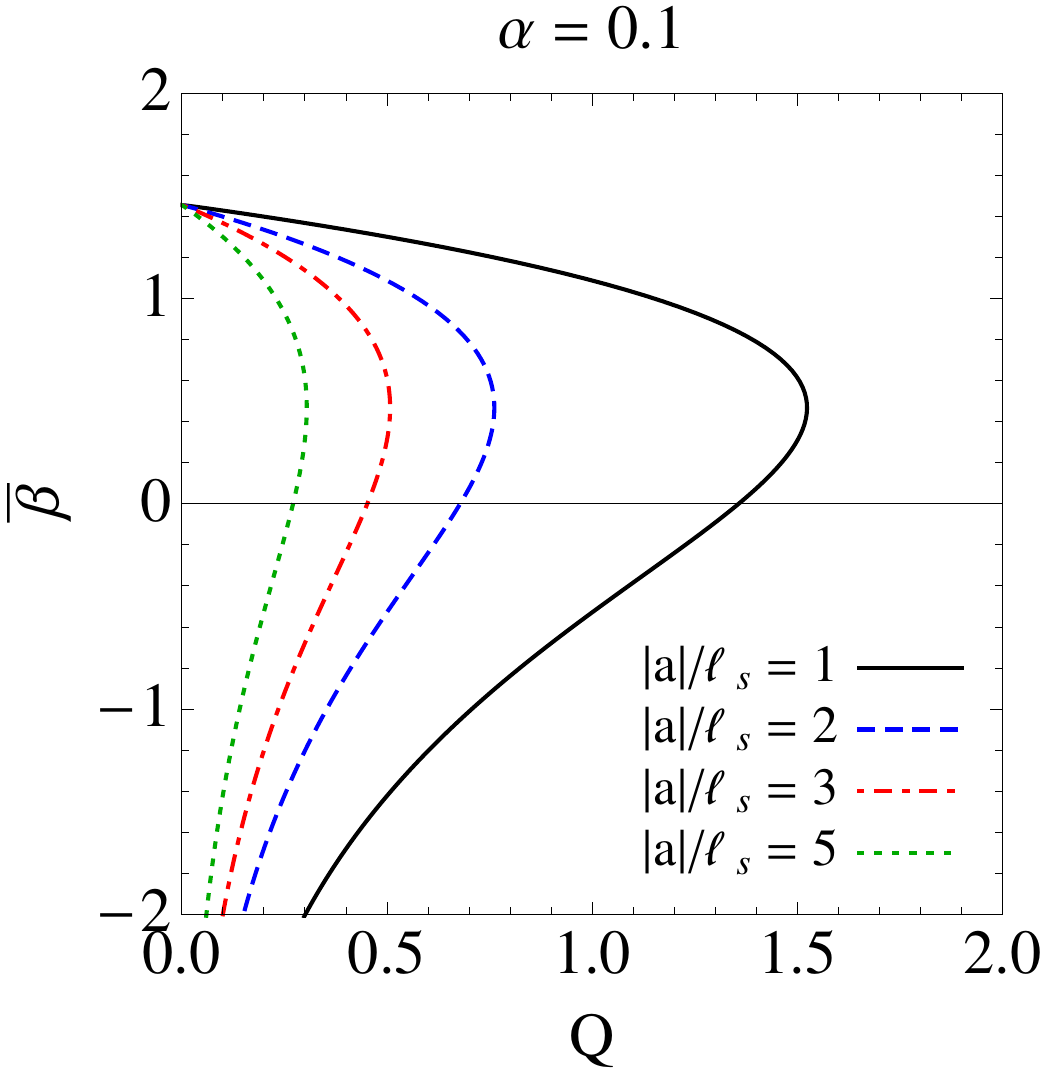} 
 \end{subfigure} \hspace{0.3cm}
  \begin{subfigure}{}
  \includegraphics[width=0.28\textwidth]{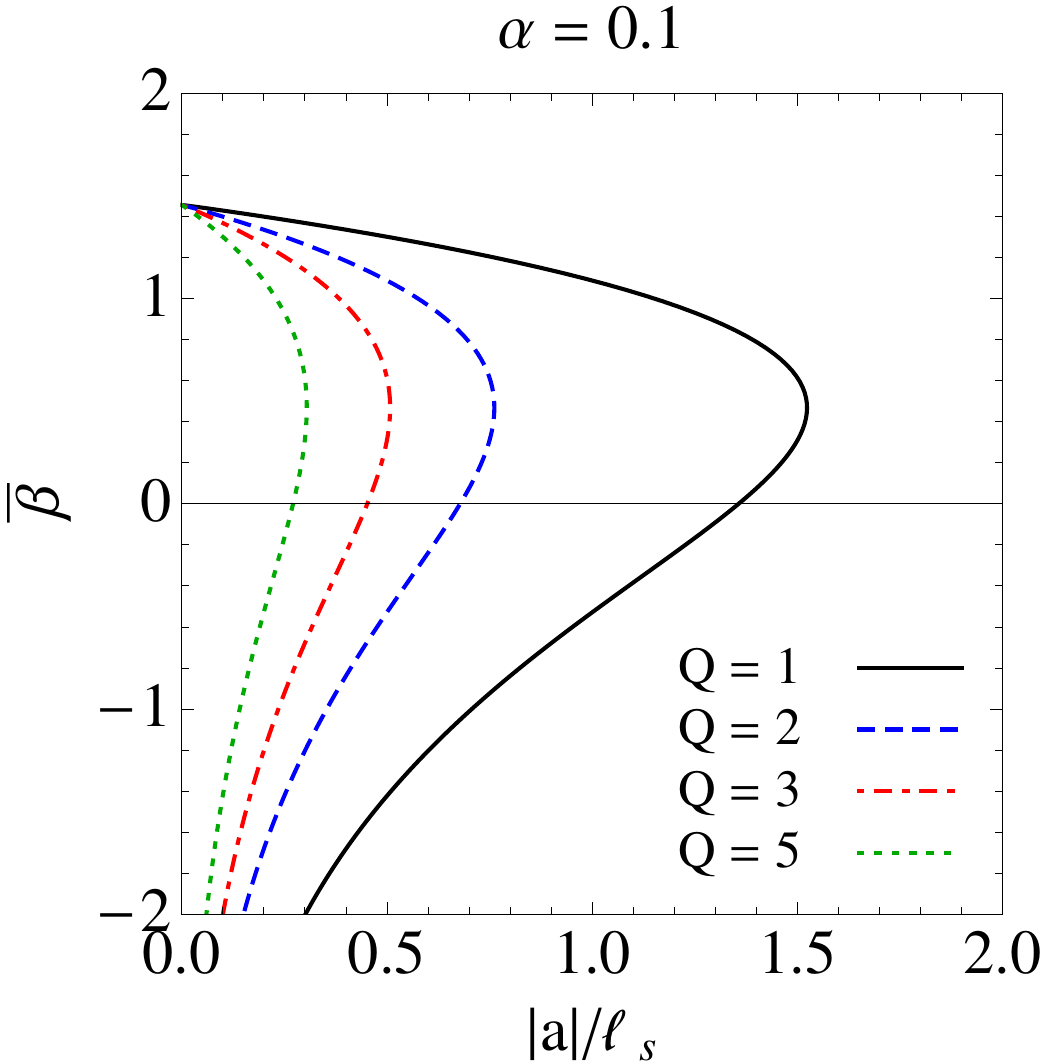} 
 \end{subfigure} \hspace{0.3cm}
 \begin{subfigure}{}
  \includegraphics[width=0.28\textwidth]{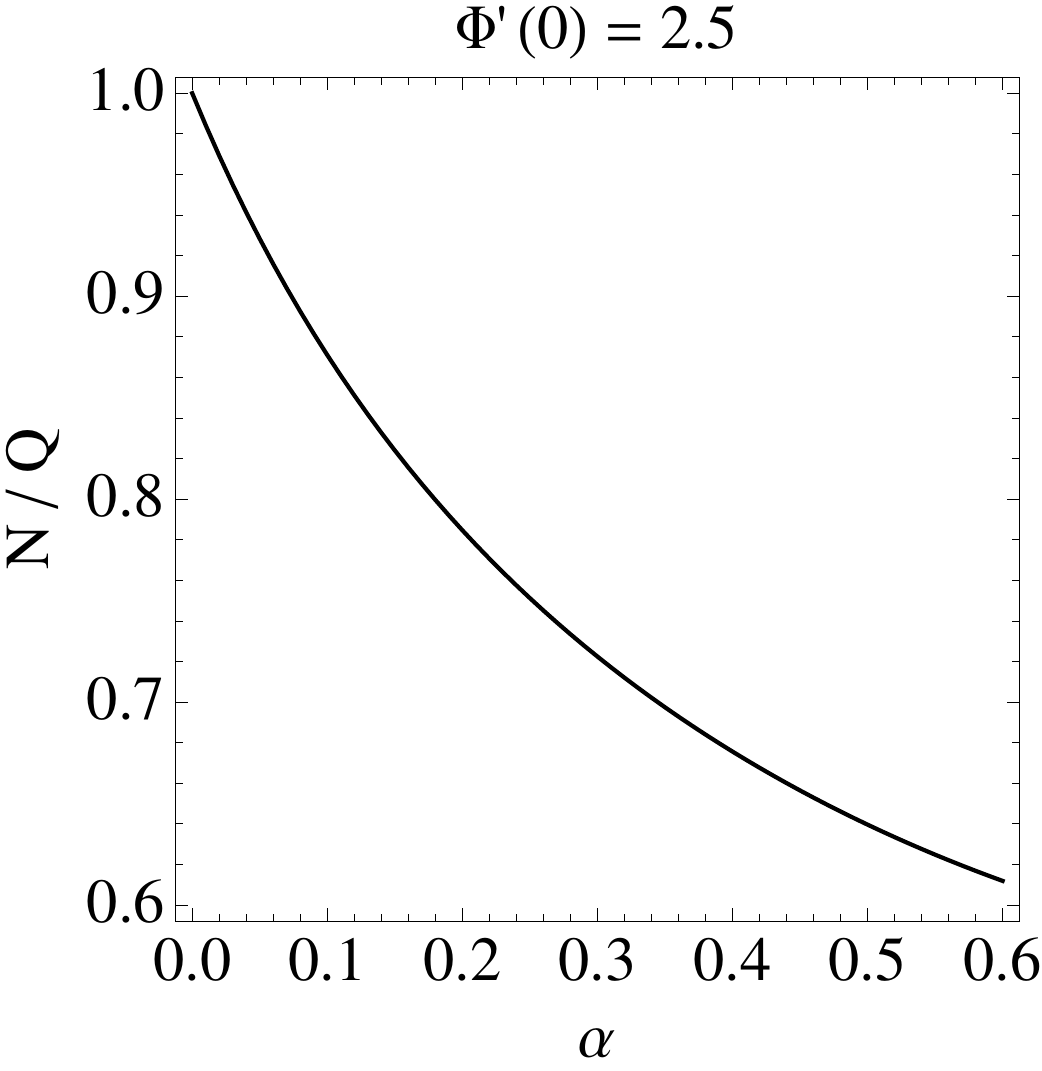} 
 \end{subfigure}
 \caption{Left and middle panels: Dimensionless chemical potential $\bar\beta$ as a function of $Q$ for fixed values of the scattering length: $|a|/\ell_s = 1,2,3$ and $5$ (left panel), and as a function of $|a|/\ell_s$ for fixed values of $Q$: $Q = 1,2,3$ and $5$ (middle panel) (cf. Eq.~(\ref{eq:def_qPhi})). We have considered $\alpha = 0.1$ in both panels. Right panel: $N/Q$ as a function of $\alpha$ for $\Phi^\prime(0) = 2.5$.
 }
 \label{fig:Pitaevskii_muQ}
\end{figure}

Finally, let us note that the eigenvalue $\mu_n^{\KG}$ is real only for real values of $\beta_{+}$, and this happens for
\begin{equation}
    g < \frac{\hbar c}{2} \,,
\end{equation}
i.e. for $\lambda < 1/2$. One can see that in the regime $g > \hbar c/2$, the asymptotic behaviors of the reduced wave function are
\begin{equation}
    u(r) \stackrel[r \to 0]{\sim}{} r^{\frac{1}{2} + i \textrm{Im}(\beta_+)} \,, \qquad\qquad u(r) \stackrel[r \to \infty]{\sim}{} e^{-\kappa r}  \qquad  \textrm{with} \qquad \textrm{Re}(\kappa) > 0 \,,
\end{equation}
so that while it is a square-integrable function, it is not a regular solution at $r \to 0$.

~

\noindent{\it System of $N$ interacting particles:} Another interesting case that we will study is a system of $N$ interacting particles under the influence of an external potential. In the non-relativistic limit, such a system is usually described in terms of the Gross-Pitaevskii nonlinear Schr\"odinger equation, which is given by
\begin{equation}
    -\frac{\hbar^2}{2m} \vec{\nabla}^2 \psi(\vec{r}) + V(\vec{r}) \psi(\vec{r}) = \bar{\mu} \psi(\vec{r}) \,. \label{eq:GP_1}
\end{equation}
In Ref.~\cite{Gammal:1999} (see also \cite{Adhikari:2000}) this system was studied with an harmonic external potential and a mean-field potential for the interaction term given in terms of a scattering-length, $a$, that is,
\begin{equation}
V(r) = \frac{1}{2} m \omega^2 r^2 - \frac{4\pi \hbar^2 |a|}{m} |\psi(r)|^2 \,.    \label{eq:GP_2}
\end{equation}
Here we study this case in the relativistic regime, and evaluate its non-relativistic limit comparing our results with the results from Ref.~\cite{Gammal:1999}. In the relativistic case, from Eq.~(\ref{eq:u}) and using the potential of Eq.~(\ref{eq:GP_2}), one has
\begin{equation}
   u^{\prime\prime}(r)  + \frac{1}{\hbar^2 c^2}\left[ \left(\mu - \frac{1}{2} m \omega^2 r^2    +  \frac{4\pi \hbar^2 |a|}{m} \frac{|u(r)|^2}{r^2} \right)^2 - m^2 c^4 \right] u(r) = 0 \,.  \label{eq:KG_Pitaevskii}  
\end{equation}
Let us define the dimensionless variables
\begin{equation}
x \equiv \sqrt{\frac{2m\omega}{\hbar}} \cdot r \,, \quad \Phi(x) \equiv \sqrt{8\pi |a|} \cdot u(r) \,, \quad \beta \equiv \frac{\mu}{\hbar \omega} \,, \quad \bar\beta \equiv \frac{\bar\mu}{\hbar\omega} \,, \quad \alpha \equiv \frac{\hbar\omega}{mc^2} \,.  \label{eq:dim_var_GP}   
\end{equation}
Using Eq.~(\ref{eq:mu_mub}) one finds $\beta = 1/\alpha + \bar\beta$. Then, Eq.~(\ref{eq:KG_Pitaevskii}) can be written in the equivalent form
\begin{equation}
       \Phi^{\prime\prime}(x)  + \left( \bar\beta - \frac{1}{4} x^2 +  \frac{|\Phi(x)|^2}{x^2} \right)  \left[ 1 + \frac{\alpha}{2} \left( \bar\beta - \frac{1}{4} x^2 +  \frac{|\Phi(x)|^2}{x^2} \right) \right] \Phi(x) = 0 \,.  \label{eq:KG_Pitaevskii2}  
\end{equation}
The non-relativistic limit corresponds to $\alpha \to 0$, or equivalently $c \to \infty$. In the non-relativistic case, we define the normalization as $N = \int d^3r |\psi(r)|^2 = 4\pi \int_0^\infty dr \, r^2 |u(r)|^2$ where $N$ is the number of atoms or partons. The corresponding normalization for the dimensionless wave function is
\begin{equation}
    n = \int_0^\infty dx \, |\Phi(x)|^2 \,, \qquad \textrm{where} \qquad n \equiv N \frac{|a|}{\ell_s}  \quad \textrm{and} \quad  \ell_s \equiv \frac{1}{2} \sqrt{ \frac{\hbar}{2m\omega} }  \,. \label{eq:def_nPhi}
\end{equation}
The length $\ell_s$ is the characteristic length of the system. In the relativistic case, it is better to define the normalization in terms of the conserved charge, i.e.
\begin{equation}
    Q = 4\pi \int_0^\infty dr \, r^2 \rho(r) \,, \qquad   \textrm{with} \qquad \rho(r) = |\varphi(r)|^2 - |\chi(r)|^2 \,, \label{eq:def_QPhi}
\end{equation}
which for the dimensionless spinor components of the wave function $\Phi_{f}(x) \equiv \sqrt{8\pi |a|} \cdot r f(r) \,, \; f = \varphi, \chi$, leads to
\begin{equation}
    q = \int_0^\infty dx \, \left( |\Phi_\varphi(x)|^2 - |\Phi_\chi(x)|^2 \right) \,, \qquad \textrm{where} \qquad q \equiv Q \frac{|a|}{\ell_s}  \,. \label{eq:def_qPhi}
\end{equation}
Obviously $Q \stackrel[c \to \infty]{\longrightarrow}{} N$ and $q \stackrel[c \to \infty]{ \longrightarrow}{} n$. We show in Fig.~\ref{fig:Pitaevskii_rel} the results of $\bar\beta$ and $q$ by solving the Gross-Pitaevskii nonlinear equation in the relativistic case, Eq.~(\ref{eq:KG_Pitaevskii2}), and compared with the non-relativistic result $(\alpha=0)$ of Ref.~\cite{Gammal:1999}. Even though the curves of Fig.~\ref{fig:Pitaevskii_rel} contain all the relevant information, it would be interesting to show explicitly the dependence of $\bar\beta$ with the scattering length~$a$, by using that $\bar\beta(q) = \bar\beta\left( Q |a|/\ell_s\right)$. To do so, we display in Fig.~\ref{fig:Pitaevskii_muQ} the results of $\bar\beta$ as a function of $Q$ for several values of $|a|/\ell_s$ (left panel), and as a function of $|a|/\ell_s$ for several values of $Q$ (middle panel). One can see that for fixed~$\bar\beta$, the larger the value of $|a|$, the smaller the value of $Q$. This can be easily understood by noticing that the relation $\bar\beta = \bar\beta(q)$ given by Fig.~\ref{fig:Pitaevskii_rel} (middle) can be inverted to give the one-valued function $q = q(\bar\beta)$, so that using Eq.~(\ref{eq:def_qPhi}) one finds $Q = \ell_s q(\bar\beta)/|a|$. Notice that the normalization of $\bar\beta$ given by Eq.~(\ref{eq:dim_var_GP}) doesn't contain any dependence in $a$, so that the dependence of $\bar\mu$ with $a$ is proportional to the dependence of $\bar\beta$ with $a$, i.e. $\bar\mu(a) = \hbar \omega \cdot \bar\beta\left(Q |a|/\ell_s \right)$. Finally, we display in Fig.~\ref{fig:Pitaevskii_muQ} (right) the dependence of $N/Q$ with the parameter~$\alpha$ for $\Phi^\prime(0) = 2.5$, where $N$ is computed in the relativistic case by using Eq.~(\ref{eq:def_nPhi}) with $\Phi(x) = \Phi_\varphi(x) + \Phi_\chi(x)$ the solution of Eq.~(\ref{eq:KG_Pitaevskii2}). Notice that in the non-relativistic limit $(\alpha \to 0)$ both quantities are equal, but in the relativistic regime one finds $N < Q$.

~

\noindent{\it Discussion and conclusions: } Concluding, in this work we analysed the time independent solutions of the non-linear Klein-Gordon equation. We used the Feshbach-Villars approach with potential, and analysed the behaviour of the system under different external potentials, namely, the Coulombic potential and the harmonic potential. 

With the introduction of the chemical potential as the eigenvalue of the Hamiltonian operator for the ground-state, we obtained the relevant conditions for the formation of the Bose-Einstein condensate in a relativistic system. The use of the Feshbach-Villars formalism allowed us to easily investigate the transition of the system from the relativistic to the non-relativistic regimes. We compare the results obtained for the relativistic system with a mean-field potential and an harmonic external potential, with those for the non-relativistic system under the same conditions. We observe a smooth transition of several parameters as the relativistic effects diminish, for both the Coulombic potential and the harmonic potential.

The effects of the interaction between the particles of the system, described by the scattering length in a mean-field approach, is analysed, and in particular, we study the variation of the chemical potential with the scattering length.

The study of the nonlinear Klein-Gordon equation to investigate condensation of relativistic particles invites us to pursue extensions of this work. One possibility is to use the same formalism of the present work to investigate condensation in the relativistic hadronic matter, and another option is to study a modified form of the Gross-Pitaevskii equation with fractional derivatives.

~

\noindent{\bf Acknowledgments} The work of E M is supported by the project PID2020-114767GB-I00 financed by MCIN/AEI/10.13039/ 501100011033, by the FEDER/Junta de Andaluc\'{\i}a-Consejer\'{\i}a de Econom\'{\i}a y Conocimiento 2014-2020 Operational Program under Grant A-FQM-178-UGR18, by Junta de Andaluc\'{\i}a under Grant FQM-225, and by the Consejería de Conocimiento, Investigaci\'on y Universidad of the Junta de Andaluc\'{\i}a and European Regional Development Fund (ERDF) under Grant SOMM17/6105/ UGR. The research of E M is also supported by the Ram\'on y Cajal Program of the Spanish MCIN under Grant RYC-2016-20678. M J T is supported by FAPESP grant 2021/12954-5. V S T is supported by FAEPEX (grant 3258/19), FAPESP (grant 2019/010889-1) and CNPq (grant 306615/2018-5). A D is partially supported by the Conselho Nacional de Desenvolvimento Cient\'{\i}fico e Tecnol\'ogico (CNPq-Brazil), grant 304244/2018-0, by Project INCT-FNA Proc. No. 464 898/2014-5 and  supported by FAPESP grant 2016/17612-7. 

\vspace{0.5cm}

\noindent{\bf Funding} Funding for open access charge: Universidad de Granada / CBUA.

\vspace{0.5cm}

\noindent{\bf Data Availability Statement} Data sharing not applicable to this article as no datasets were generated or analysed during the current study.

%
 \bibliographystyle{elsarticle-num}
 \bibliography{bib2.bib}

\begin{thebibliography}{10}
\expandafter\ifx\csname url\endcsname\relax
  \def\url#1{\texttt{#1}}\fi
\expandafter\ifx\csname urlprefix\endcsname\relax\def\urlprefix{URL }\fi
\expandafter\ifx\csname href\endcsname\relax
  \def\href#1#2{#2} \def\path#1{#1}\fi

\bibitem{Hogan}
C.~J. Hogan, M.~J. Rees, Axion miniclusters, Physics Letters B 205~(2-3) (1988)
  228--230.

\bibitem{Frieberg}
R.~Friedberg, T.~D. Lee, Y.~Pang, Scalar soliton stars and black holes,
  Physical Review D 35~(12) (1987) 3658.

\bibitem{EbyLeembruggen}
J.~Eby, M.~Leembruggen, L.~Street, P.~Suranyi, L.~Wijewardhana, {Global view of
  QCD axion stars}, Physical Review D 100~(6) (2019) 063002.

\bibitem{EbyMadelyn}
J.~Eby, M.~Leembruggen, P.~Suranyi, L.~Wijewardhana, Stability of condensed
  fuzzy dark matter halos, Journal of Cosmology and Astroparticle Physics
  2018~(10) (2018) 058.

\bibitem{GrundlandInfeld}
A.~M. Grundland, E.~Infeld, {A family of nonlinear Klein--Gordon equations and
  their solutions}, Journal of Mathematical Physics 33~(7) (1992) 2498--2503.

\bibitem{EbyStreet}
J.~Eby, M.~Leembruggen, L.~Street, P.~Suranyi, L.~Wijewardhana, Approximation
  methods in the study of boson stars, Physical Review D 98~(12) (2018) 123013.

\bibitem{EbySuranyi}
J.~Eby, P.~Suranyi, L.~Wijewardhana, {Expansion in higher harmonics of boson
  stars using a generalized Ruffini-Bonazzola approach. Part 1. Bound states},
  Journal of Cosmology and Astroparticle Physics 2018~(04) (2018) 038.

\bibitem{Rezaeian}
A.~H. Rezaeian, {Semi-inclusive photon-hadron production in \emph{pp} and
  \emph{pA} collisions at RHIC and LHC}, Physical Review D 86~(9) (2012)
  094016.

\bibitem{Jankowski}
J.~Jankowski, D.~Blaschke, M.~Spali{\'n}ski, Chiral condensate in hadronic
  matter, Physical Review D 87~(10) (2013) 105018.

\bibitem{Fujii}
H.~Fujii, Y.~Tsue, {Quark condensate in a medium and pion-nucleon sigma term
  based on a QCD-like theory}, Physics Letters B 357~(1-2) (1995) 199--203.

\bibitem{Ketterle-vanDruten-PRL-1996}
M.~O. Mewes, M.~R. Andrews, N.~J. van Druten, D.~M. Kurn, D.~S. Durfee, C.~G.
  Townsend, W.~Ketterle, {Collective excitations of a Bose-Einstein condensate
  in a magnetic trap}, Physical Review Letters 77~(6) (1996) 988--991.
\newblock \href {https://doi.org/10.1103/PhysRevLett.77.988}
  {\path{doi:10.1103/PhysRevLett.77.988}}.

\bibitem{Bagnato-Pritchard-Kleppner}
V.~Bagnato, D.~E. Pritchard, D.~Kleppner, {Bose-Einstein condensation in an
  external potential}, Physical Review A 35~(10) (1987) 4354--4358.
\newblock \href {https://doi.org/10.1103/PhysRevA.35.4354}
  {\path{doi:10.1103/PhysRevA.35.4354}}.

\bibitem{RepulsiveInteractionBEC}
O.~Savchuk, Y.~Bondar, O.~Stashko, R.~V. Poberezhnyuk, V.~Vovchenko, M.~I.
  Gorenstein, H.~Stoecker, {Bose-Einstein condensation phenomenology in systems
  with repulsive interactions}, Phys. Rev. C 102 (2020) 035202.
\newblock \href {https://doi.org/10.1103/PhysRevC.102.035202}
  {\path{doi:10.1103/PhysRevC.102.035202}}.

\bibitem{Plastino-Tsallis-2016}
A.~R. Plastino, C.~Tsallis, {Dissipative effects in nonlinear Klein-Gordon
  dynamics}, {EPL} (Europhysics Letters) 113~(5) (2016) 50005.
\newblock \href {https://doi.org/10.1209/0295-5075/113/50005}
  {\path{doi:10.1209/0295-5075/113/50005}}.

\bibitem{Guerrero-Gonzalez}
L.~E. Guerrero, J.~A. Gonzalez, {Long-range interacting solitons: pattern
  formation and nonextensive thermostatistics}, Physica A 257~(1-4) (1998)
  390--394, vth Latin American Workshop on Non-Linear Phenomena/11th MEDYFINOL
  Conference on Statistical Physics of Dynamic and Complex Systems, Canela,
  Brazil, Sep 28 -- Oct 03, 1997.
\newblock \href {https://doi.org/10.1016/S0378-4371(98)00165-4}
  {\path{doi:10.1016/S0378-4371(98)00165-4}}.

\bibitem{Megias-Varese-Gammal-Deppman-2021}
E.~Megías, V.~Timóteo, A.~Gammal, A.~Deppman, {Bose–Einstein condensation
  and non-extensive statistics for finite systems}, Physica A: Statistical
  Mechanics and its Applications 585 (2022) 126440.
\newblock \href {https://doi.org/https://doi.org/10.1016/j.physa.2021.126440}
  {\path{doi:https://doi.org/10.1016/j.physa.2021.126440}}.

\bibitem{Rajagopal-Lenzi}
A.~K. Rajagopal, R.~S. Mendes, E.~K. Lenzi, {Quantum statistical mechanics for
  nonextensive systems: Prediction for possible experimental tests}, Physical
  Review Letters 80~(18) (1998) 3907--3910.

\bibitem{Gammal:1999}
A.~Gammal, T.~Frederico, L.~Tomio, {Improved numerical approach for the
  time-independent Gross-Pitaevskii nonlinear Schr{\"o}dinger equation},
  Physical Review E 60~(2) (1999) 2421.

\bibitem{Adhikari:2000}
S.~K. Adhikari, {Numerical study of the spherically symmetric Gross-Pitaevskii
  equation in two space dimensions}, Physical Review E 62~(2) (2000)
  2937--2944.

\bibitem{Feshbach:1958wv}
H.~Feshbach, F.~Villars, Elementary relativistic wave mechanics of spin 0 and
  spin 1/2 particles, Reviews of Modern Physics 30~(1) (1958) 24.

\bibitem{Alberto:2017pkj}
P.~Alberto, S.~Das, E.~C. Vagenas, {Relativistic particle in a box:
  Klein–Gordon versus Dirac equations}, Eur. J. Phys. 39~(2) (2018) 025401.
\newblock \href {https://doi.org/10.1088/1361-6404/aa9b43}
  {\path{doi:10.1088/1361-6404/aa9b43}}.

\bibitem{Dominguez-Adame:1989gml}
F.~Dominguez-Adame, {Bound states of the {Klein-Gordon} equation with vector
  and scalar Hulthen type potentials}, Phys. Lett. A 136 (1989) 175--177.
\newblock \href {https://doi.org/10.1016/0375-9601(89)90555-0}
  {\path{doi:10.1016/0375-9601(89)90555-0}}.

\bibitem{Chen:2005}
G.~Chen, Z.~D. Chen, {Bound stated solutions of the Klein-Gordon equation with
  Hartmann potential and recursion relations}, Acta Phys. Sin. 54~(6) (2005)
  2524--2527.

\bibitem{Zhang:2005}
X.~A. Zhang, K.~Chen, Z.~L. Duan, {{Bound States of the Klein-Gordon equation
  and Dirac equation for ring-shaped non-spherical oscillator scalar and vector
  potentials}}, Chin. Phys. 14~(1) (2005) 42--44.

\bibitem{Chen:2005plA}
C.~Y. Chen, {Exact solutions of the Dirac equation with scalar and vector
  Hartmann potentials}, Phys. Lett. A 339 (2005) 283--287.

\bibitem{ChenLu:2006}
C.~Y. Chen, F.~L. Lu, D.~S. Sun, {The relativistic bound states of Coulomb
  potential plus a new ring-shaped potential}, Acta Phys. Sin. 55~(8) (2006)
  3875--3879.

\bibitem{Zhang:2006}
M.~C. Zhang, Z.~B. Wang, {Bound states of the {Klein-Gordon} equation and Dirac
  equation with the Manning-Rosen scalar and vector potentials}, Acta Phys.
  Sin. 55~(2) (2006) 521--524.

\bibitem{Ducharme:2010}
R.~J. Ducharme, {Exact solution of the Klein-Gordon equation for the hydrogen
  atom including electron spin} (2010).
\newblock \href {http://arxiv.org/abs/1006.3971} {\path{arXiv:1006.3971}}.

\bibitem{Sharma:2011}
L.~K. Sharma, P.~V. Luhanga, S.~Chimidza, {Potentials for the Klein-Gordon and
  Dirac equations}, Chiang Mai Journal of Science 38~(4) (2011) 514--526.

\bibitem{Bi:2012rx}
G.~Bi, Y.~Bi, {New relativistic wave equations for two-particle systems} (4
  2012).
\newblock \href {http://arxiv.org/abs/1204.4876} {\path{arXiv:1204.4876}}.

\bibitem{Silenko:2019brp}
A.~J. Silenko, {Zitterbewegung of bosons}, Phys. Part. Nucl. Lett. 17~(2)
  (2020) 116--119.
\newblock \href {https://doi.org/10.1134/S1547477120020193}
  {\path{doi:10.1134/S1547477120020193}}.

\end{thebibliography}
%
%
%

\end{document}